# Valid Physical Processes from Numerical Discontinuities in Computational Fluid Dynamics

Kun Xu[1*], Quanhua Sun[2], and Pubing Yu[1]

[1]*Department of Mathematics, The Hong Kong University of Science and Technology, Kowloon, Hong Kong*
[2]*LHD, Institute of Mechanics, Chinese Academy of Sciences, Beijing, 100190, China*

Due to the limited cell resolution in the representation of flow variables, a piecewise continuous initial reconstruction with discontinuous jump at a cell interface is usually used in modern computational fluid dynamics methods. Starting from the discontinuity, a Riemann problem in the Godunov method is solved for the flux evaluation across the cell interface in a finite volume scheme. With the increasing of Mach number in the CFD simulations, the adaptation of the Riemann solver seems introduce intrinsically a mechanism to develop instabilities in strong shock regions. Theoretically, the Riemann solution of the Euler equations are based on the equilibrium assumption, which may not be valid in the non-equilibrium shock layer. In order to clarify the flow physics from a discontinuity, the unsteady flow behavior of one-dimensional contact and shock wave is studied on a time scale of (0~10000) times of the particle collision time. In the study of the non-equilibrium flow behavior from a discontinuity, the collision-less Boltzmann equation is first used for the time scale within one particle collision time, then the direct simulation Monte Carlo (DSMC) method will be adapted to get the further evolution solution. The transition from the free particle transport to the dissipative Navier-Stokes (NS) solutions are obtained as an increasing of time. The exact Riemann solution becomes a limiting solution with infinite number of particle collisions. For the high Mach number flow simulations, the points in the shock transition region, even though the region is enlarged numerically to the mesh size, should be considered as the points inside a highly non-equilibrium shock layer. In order to develop a robust and accurate numerical scheme for all speed flows, the physical processes from the non-equilibrium to equilibrium with the increasing of particle collisions need to be followed. The use of exact Riemann solution in the Godunov method lacks the mechanism to describe the non-equilibrium flow behavior from a discontinuity. On the other hand, the gas-kinetic scheme (GKS) follows the non-equilibrium flow physics and its evolution to equilibrium state, which may be the reason for its absence of shock instabilities in high Mach number flow computations. At the same, the accurate NS solutions can be obtained in the smooth flow regions, such as the boundary layer, due to its quick adaptation to the equilibrium states in these regions.

* corresponding author. Phone: 852-23587433, Email: makxu@ust.hk

# I. INTRODUCTION

The Boltzmann equation is generally regarded as the governing equation for the motion of fluid. It describes the time evolution of a large number of particles through binary collisions in statistical physics. This is a seven-dimensional integral-differential equation, which is more fundamental than the Euler and Navier-Stokes equations. This equation, however, can be simplified under some conditions. For the equilibrium fluid, the Boltzmann equation leads to the compressible Euler system which is a nonlinear hyperbolic system of conservation laws. The basic wave structure of the hyperbolic system, such as shock wave, contact discontinuity, and rarefaction wave, has been well studied in the past decades. Among these waves, shock wave and contact discontinuity are considered to be simple jump discontinuities, and how to capture them numerically motivated the development of modern CFD methods [1,2]. Even though great success has obtained achieved in CFD for the compressible flow computations in the past decades, for high Mach number flow computation most numerical schemes based on the Euler solution from a discontinuity, i.e., the Riemann solution, have encountered great numerical difficulties, such as the emerging of shock instabilities or the carbuncle phenomena [3,4]. The exact Riemann solution is based on the equilibrium flow solution everywhere connected through discontinuities. For high Mach number flows, when the numerical cell size is on the same order of particle mean free path $l$, such as inside the numerical shock layer, the development of a valid CFD algorithm under this situation has to take into account the non-equilibrium flow behavior associated with particle transport and collisions. The non-equilibrium flow behavior converges to the equilibrium one only in the case with massive particle collisions. In general, it may be necessary to use the non-equilibrium flow physics to follow up the time evolution of initial discontinuities in the design reliable numerical schemes.

When the flow deviates slightly from local equilibrium in the continuum flow regime, the approximate solution derived from the Boltzmann equation leads to the Navier-Stokes equations [5], where the dissipative terms being proportional to the gradients of velocity and temperature. In this case, shock wave and contact discontinuity are not mathematically discontinuities anymore due to viscous diffusion and heat conductivity. In most times, the physical thickness of shock wave and contact discontinuity can be much smaller than the mesh size, which can be treated theoretically as jump discontinuities when solving the Navier-Stokes equations. However, the numerical shock thickness for any shock capturing scheme should always be on the order of mesh size. To correctly capture an enlarged shock thickness, the non-equilibrium physics may be needed in the numerical shock layer instead of equilibrium Euler solution locally.

In this paper, we will first present the numerical difficulties in modern CFD methods which are based on the Riemann solution. In section 3, we are going to present the real flow physics from a discontinuity. More specifically, we will study the unsteady behavior of both shock wave and contact discontinuity from a simple mathematical jump to the development of a well-defined dissipative structure. Then, in section 4, following the non-equilibrium flow evolution process, we will present the methodology of the gas-kinetic scheme (GKS). A few numerical examples from the GKS will be presented to illustrate its robustness and accuracy in the high speed Mach number flow computations. The last section is the conclusion which states the valid physical process which should be used in the development of modern CFD methods.

## II. NUMERICAL DIFFICULTIES FOR GODUNOV-TYPE SCHEMES

The modern CFD method for compressible flow is based on the Riemann problem from a piecewise constant states [6,7]. The necessity to use discontinuous initial condition is due to the limited cell resolution to represent physical flow structure. Due to the preparation of discontinuous initial data through the so-called nonlinear limiter, the numerical dissipation is implicitly added in the shock capturing schemes [8]. In the past decades, the shock capturing CFD methods based on the exact or approximate Riemann problems are extremely successful in the aerospace engineering applications and the scientific study of compressible flows. However, when going to hypersonic flow computation, i.e., M>10, the controversy between accuracy and robustness of a numerical scheme appears. A outstanding challenge is the shock instability and carbuncle phenomenon in the blunt body simulations [3,9], see Fig. 1. It seems that for high Mach number flow most Riemann solvers are intrinsically rooted with the shock instability, except for a few very dissipative schemes. Many have proposed cures to the shock instability, but none are universally accepted at the current stage. It is well-known that adding numerical viscosity in the fluxes could prevent these problems but with an unavoidable lost of accuracy. There are proposals to adopt a hybrid of very dissipative and less dissipative fluxes, deploying the former near the shock and the latter away from shock, but the basis of the switch is somewhat ad hoc. Furthermore, it is not clear how any switch would work for complex problems like shock-boundary layer interactions or shock-contact interactions. If the physical viscosity is included as in the Navier-Stokes equations, the tendency to form a carbuncle is reduced, but it disappears only at very low Reynolds number, where the physical dissipation suppresses the numerical instability.

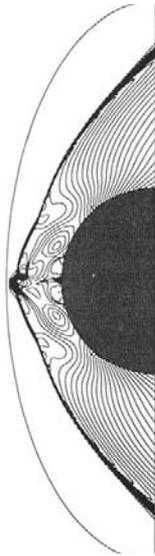

FIG. 1. Shock instability. Density contours around blunt body at M=10 (James Quirk, Int. Journal for Numerical Methods in Fluids, Vol. 18, pp. 555-574 (1994)).

The failure of the shock capturing scheme may be due to the inconsistent treatment of flow physics from a discontinuity. The Riemann solution assumes equilibrium flow everywhere. However, inside the numerical shock layer, theoretically the numerical cell size is on the same order of numerically enlarged particle mean free path. Therefore, on the scale of numerical cell size, the non-equilibrium flow physics must be taken into account in the gas evolution from a discontinuity. Any point between the upstream and downstream shock condition needs to be considered as a point inside a non-equilibrium shock structure. A valid physical process from a discontinuity initial condition, at least for the gas, should be the one from collision-less transport to the formation of equilibrium state due to particle collisions, then the drifting of equilibrium states to generate the dissipative NS solution. The Euler solution in the Riemann problem replaces the above non-equilibrium evolution process by an equilibrium solution which assumes infinite number of particle collisions anywhere. It may be true in smooth flow region. In the region with high gradients, the particle transport from one state to another stage may not take infinite number of collisions at all. The mis-representation of a non-equilibrium shock structure by the Euler Riemann solution can be also understood in the following example. Suppose that there is a very strong shock wave with upstream and down stream shock condition $(W_u, W_d)$, and these two states can be connected through a Rankine-Hugoniot condition. For a numerical scheme, due to the conservation and averaging it is most likely that there is a point between the above two states $W_m$. In order to get time evolution for the above shock solution, the Godunov method needs to solve two Riemann problems for the states $(W_u, W_m)$ and $(W_m, W_d)$. Since there are only three waves for the Euler solution, i.e., shock, contact discontinuity, and rarefaction waves. For a stationary strong shock, the only possible solution between states $W_u$ and $W_m$, also between $W_m$ and $W_d$, are shocks. Therefore, based on the Euler equations, the Godunov method represents a strong shock numerically by two subsequent strong shocks. However, this picture is problematic. For a single strong shock, the largest density jump is 6 for a diatomic gas. If it is composed of two strong shocks, the density jump can go to 36, which is an invalid description. In this case, the real mechanism to save the Godunov method is the conservative property of the scheme and the cell averaging which translates kinetic energy into thermal one [8]. The point inside shock layer has to be considered as a point inside a non-equilibrium shock structure, which has to be described through particle transport and collisions. In the strong shock case, the use of the Euler solution inside a numerical shock layer is inappropriate. The numerical shock structure requires that the numerical cell size is on the scale of particle mean free path. Theoretically, it is impossible to get such an equilibrium solution in the shock region.

Both gas evolution models of collision-less particle transport and equilibrium Euler solutions from a discontinuous initial data have been described by two kinds of numerical schemes which have been widely used in CFD community. The collision-less limit solution corresponds to the Flux Vector Splitting (FVS) scheme, such as Steger-Warming [10], van Leer [11], Sanders-Prendergast [12], Pullin [13], Desphande [14], and many others, while the Euler solution with "intensive" particle collision goes to the Flux Difference Splitting (FDS) methods, such as Godunov [6], Roe [15], and Osher [16]. Many other schemes, such as HLL [17], AUSM [18], can be considered somehow as a hybrid method between FVS and FDS methods. So, for the strong shock waves, the FDS methods generate numerical instability, but these schemes are accurate for the viscous boundary layer calculation due to less numerical dissipation in their inviscid flux function. On the other hand, the FVS schemes are very robust, but inaccurate for the NS solutions. In order to combine the advantages of both FVS and FDS scheme, many hybrids methods have been developed. But, the hybridization is through some kinds of averaging, where a detailed governing equation which controls the "averaging" is absent. An example is Moschetta's EFMO [19] scheme, which shares the robustness of Pullin's EFM or KFVS method and the accuracy provided by Osher's FDS method. Unfortunately, in the above hybrid methods, how to control the percentage of FVS to FDS is unknown. In order to increase the robustness of the Godunov method, the traditional treatment is to modify the eigen-values, where the drawbacks associated with this correction is that the NS solution cannot be accurately captured. For example, the boundary layers are significantly broadened when using a typical value for Harten's entropy fix function, and the exact resolution of contact waves is also lost. Someone may think to use NS equations directly to cure the shock instability. But, this is not realistic, since natural viscosity is not enough to cure this flaw by itself. Once there is a discontinuity at the cell interface, the associated particle evolution physics should be compatible with the discontinuity, such as the strength of free transport depends on the scale of flow jumps. However, for the physical viscosity term in the NS equations, there is no any consideration about the jump of the numerical discontinuity. The NS viscosity may not be enough to damp the numerical instability, especially in the high Reynolds number case.

In order to fully solve this problem, we need to follow the flow physics closely from a discontinuity. In the next section, the gas evolution from two discontinuities will be fully studied. One is the contact discontinuity and the other is the shock wave. For example, the physical shock structure is obtained through the balancing of particle transport and collisions. A highly non-equilibrium wave structure is needed in the construction of such a structure. Even though for a numerical scheme, we don't need to get the precise shock structure, but the numerical evolution process in the construction of its fluxes must be consistent with the physical one. The numerical shock structure in the shock capturing scheme can be an enlarged "physical" one. In other words, for the shock capturing schemes, even though there are only two or three transition points in the shock layer, the "numerical" shock should have a structure which has the non-equilibrium physical properties described in the next section. The trigger of carbuncle phenomena in high Mach number case is mainly due to the absence of non-equilibrium physical process in the FDS-type Riemann solution. The FVS methods have the mechanism associated with the free transport, the so-called introduction of dynamic dissipative, and the particle collision process is implicitly included in the preparation of the initial data [8], therefore the FVS schemes have a consistent physical process to construct a stable numerical shock structure and avoid carbuncle phenomena. However, this reliable shock structure is obtained through the sacrifice of accuracy, because the free transport and collision in FVS schemes use the cell size and time step as the physical mean free path and particle collision time, which could easily poison the NS solutions. For the FDS schemes, there is no a corresponding valid physical process because it models a limiting case with infinite number of particle collisions. The above analysis is consistent with the fact that the strict stability and exact resolution of contact discontinuities in a numerical scheme are not compatible. The incompatibility is due to the reason that the stability requires the free transport mechanism and the contact preservation requires the full equilibriums state. They are two limiting solutions with no particle collision and infinite number of particle collisions. In order to develop a valid scheme with both robustness and accuracy, it requires a governing equation to control the transition from FVS to FDS.

As presented in section IV, the gas-kinetic scheme (GKS) can be considered as a combination of FVS and FDS schemes as well. However, the GKS has a continuous transition from FVS to FDS, and the weighting function depends on the ratio between the time step and the particle collision time. In the GKS, the limits taken depend on the flow situation. In the dissipative shock layer, the collision-less limit plays a dominant role in the construction of non-equilibrium shock structure, and in the smooth boundary layer the NS solution limit will be obtained. Therefore, both accuracy and robustness can be kept in the GKS.

# III. VALID PHYSICAL PROCESS FROM AN INITIAL DISCONTINUITY

Mathematical discontinuities exist only in hyperbolic equations, where there is no a time scale related to the physical property of the gas. When the particle collision time appears in the mathematical modeling, the dissipative terms appear and the strong gradients of flow properties around the discontinuities will smear the discontinuities. At a time scale less than one mean collision time of the gas molecules, the flow can be predicted using the free molecular theory.

For simplicity, we assume that the original discontinuity is located at $x = 0$. Then the initial velocity distribution function $f(c,x,t)$ of the flow molecules can be expressed in one dimension as:

$$f(c,x,0) = n_1 \left(\frac{\beta_1}{\sqrt{\pi}}\right) e^{-\beta_1^2(c-u_1)^2} (1 - H(x)) + n_2 \left(\frac{\beta_2}{\sqrt{\pi}}\right) e^{-\beta_2^2(c-u_2)^2} H(x), \quad (1)$$

where $c$ is the x-component velocity of molecules, $u$ is the x-component mean flow velocity, $n$ is the number density, $\beta$ is the most probable velocity of molecules $\beta = 1/\sqrt{2RT}$, $T$ is the gas temperature, $R$ is the gas constant, $H$ is the Heaviside function, and subscripts 1 and 2 refer to the left and right sides of the discontinuity, respectively. Following the free molecular theory, the velocity distribution function at time $t$ is obtained as:

$$f(c,x,t) = n_1 \left(\frac{\beta_1}{\sqrt{\pi}}\right) e^{-\beta_1^2(c-u_1)^2} (1 - H(x-ct)) + n_2 \left(\frac{\beta_2}{\sqrt{\pi}}\right) e^{-\beta_2^2(c-u_2)^2} H(x-ct) \quad (2)$$

For contact discontinuities, we assume that the initial velocity of the discontinuities is zero. Then the distribution function can be simplified as:

$$f(c,x,t) = n_1 \left(\frac{\beta_1}{\sqrt{\pi}}\right) e^{-\beta_1^2 c^2} (1 - H(x-ct)) + n_1 \frac{T_1}{T_2} \left(\frac{\beta_2}{\sqrt{\pi}}\right) e^{-\beta_2^2 c^2} H(x-ct) \quad (3)$$

The macroscopic flow quantities can be obtained by integrating the velocity distribution function with proper quantities. They are:

$$\frac{\rho(x,t)}{\rho_1} = \frac{1}{2}\left(1 - erf\left(\beta_1 \frac{x}{t}\right)\right) + \frac{1}{2}\frac{T_1}{T_2}\left(1 + erf\left(\sqrt{\frac{T_1}{T_2}}\beta_1 \frac{x}{t}\right)\right) \quad (4)$$

$$U(x,t) = \frac{\rho_1/\rho}{2\sqrt{\pi}\beta_1}\left(e^{-\left(\beta_1 \frac{x}{t}\right)^2} - \sqrt{\frac{T_1}{T_2}}e^{-\frac{T_1}{T_2}\left(\beta_1 \frac{x}{t}\right)^2}\right) \quad (5)$$

$$T_n(x,t) = \frac{T_1}{2}\frac{\rho_1}{\rho}\left(2 + erf\left(\sqrt{\frac{T_1}{T_2}}\beta_1 \frac{x}{t}\right) - erf\left(\beta_1 \frac{x}{t}\right)\right) \quad (6)$$

$$T_x(x,t) = T_n(x,t) + \frac{U}{R}\frac{x}{t} \quad (7)$$

where $\rho$ is the mass density, $U$ is the x-component flow velocity, $T_n$ is the temperature in the direction normal to the wave propagation direction, and $T_x$ is the one in the parallel direction or x-component temperature.

From these expressions, it is clear that contact discontinuities have profiles and the width of discontinuities is proportional to the time. Details of the diffused contact discontinuities are plotted in Fig. 2, where $\lambda_1 = 1/(\sqrt{2}\sigma_1 n_1)$, $\tau_1 = \lambda_1/\sqrt{8/\pi \cdot RT}$, and $\sigma$ is the total collision cross section of argon molecules based on the VHS molecular model [20]. In the plots, the strength of the discontinuities is denoted by the temperature difference crossing the discontinuity. In general, the contact discontinuities are diffused in both sides and their width is about several molecular mean free paths at a time of one mean collision time. Stronger discontinuities diffuse slightly faster though. It is found that there appears overshoot and undershoot in the density profile (Fig. 2a), which is the result of the movement of molecules (Fig. 2b). The maximum velocity in the profile is nearly proportional to the difference of the speed of sound between two sides. The strength of discontinuities, however, determines the detailed profile of the velocity distribution. The temperature exhibits non-equilibrium behavior among the translation components. The temperature component normal to the wave propagation direction has a smooth profile, whereas the parallel component (namely, the x-component in the plot) shows complicated structure, which is due to the velocity term as shown in Eq. 7. It seems that the discontinuity in the temperature profiles shifts to the higher temperature side when the strength of the discontinuity gets stronger.

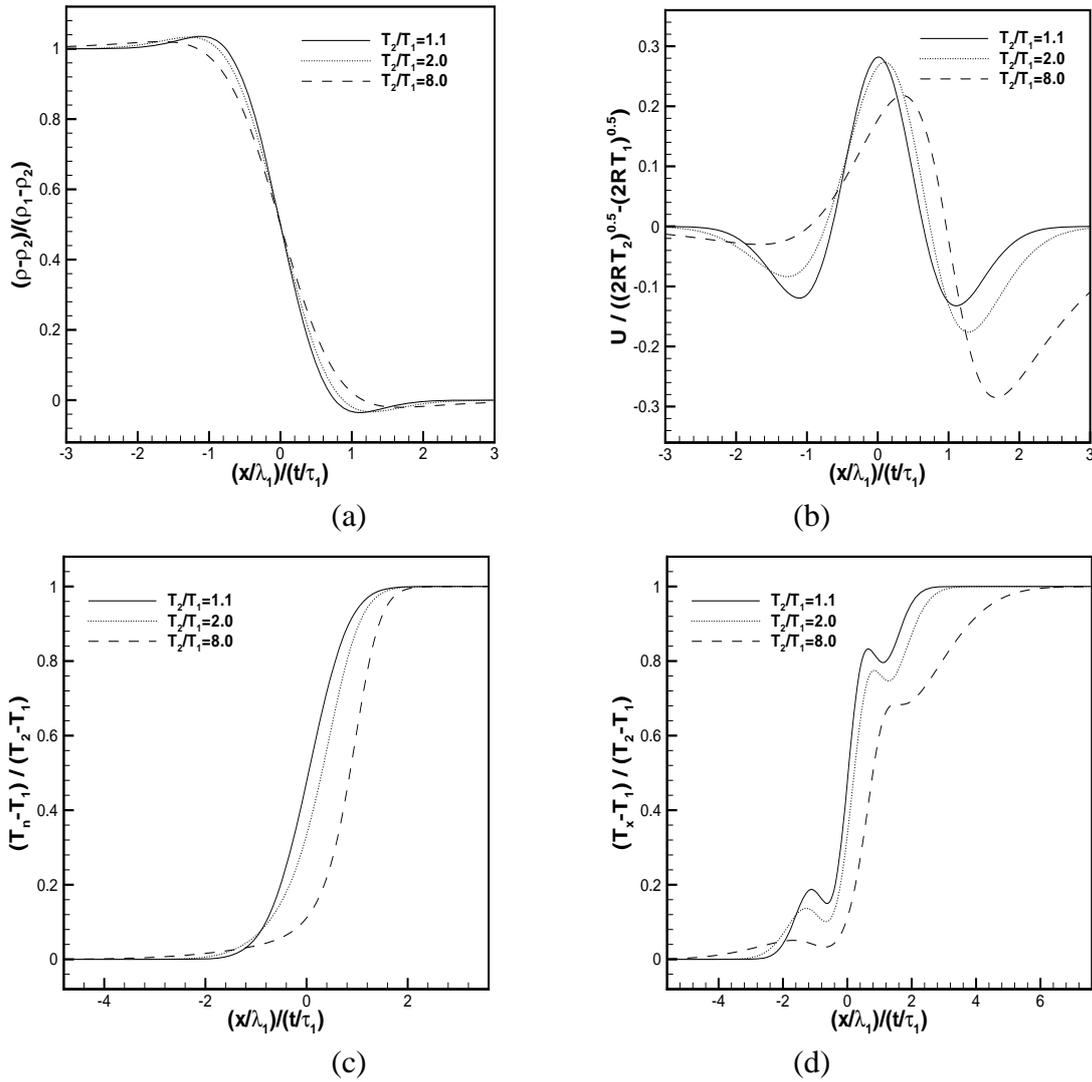

FIG. 2. Free molecular results of contact discontinuities. (a) density; (b) flow velocity; (c) normal-component temperature; (d) parallel-component temperature.

The free molecular theory is not valid when the time is larger than the mean collision time. The direct simulation Monte Carlo (DSMC) method [20] is then used to track the development of the contact discontinuity. Figure 3 shows the corresponding profiles at larger times for the case when the strength of discontinuity is 8. Notice that there are statistical scatters in the plots because unsteady simulation is numerically expensive and the minimum sampling size is only 5000 particles for our simulations. The number of simulated particles, however, does not affect the profiles except the scatters. Clearly, the width of discontinuity keeps increasing with the time. Specifically, the overshoot in the density profile increases; the velocity in the higher density side remains decreasing; for the x-component temperature, the first local maximum disappears gradually and the second local maximum increases. For even longer time (Fig. 4), the structure of the discontinuity becomes simple: the overshoot of the density profile is varnished; the temperatures among the translational components reach equilibrium, and the temperature profile looks smooth. The long time results are expected because the gradients of contact discontinuities become small due to diffusion and the flow approaches equilibrium state due to particle collisions.

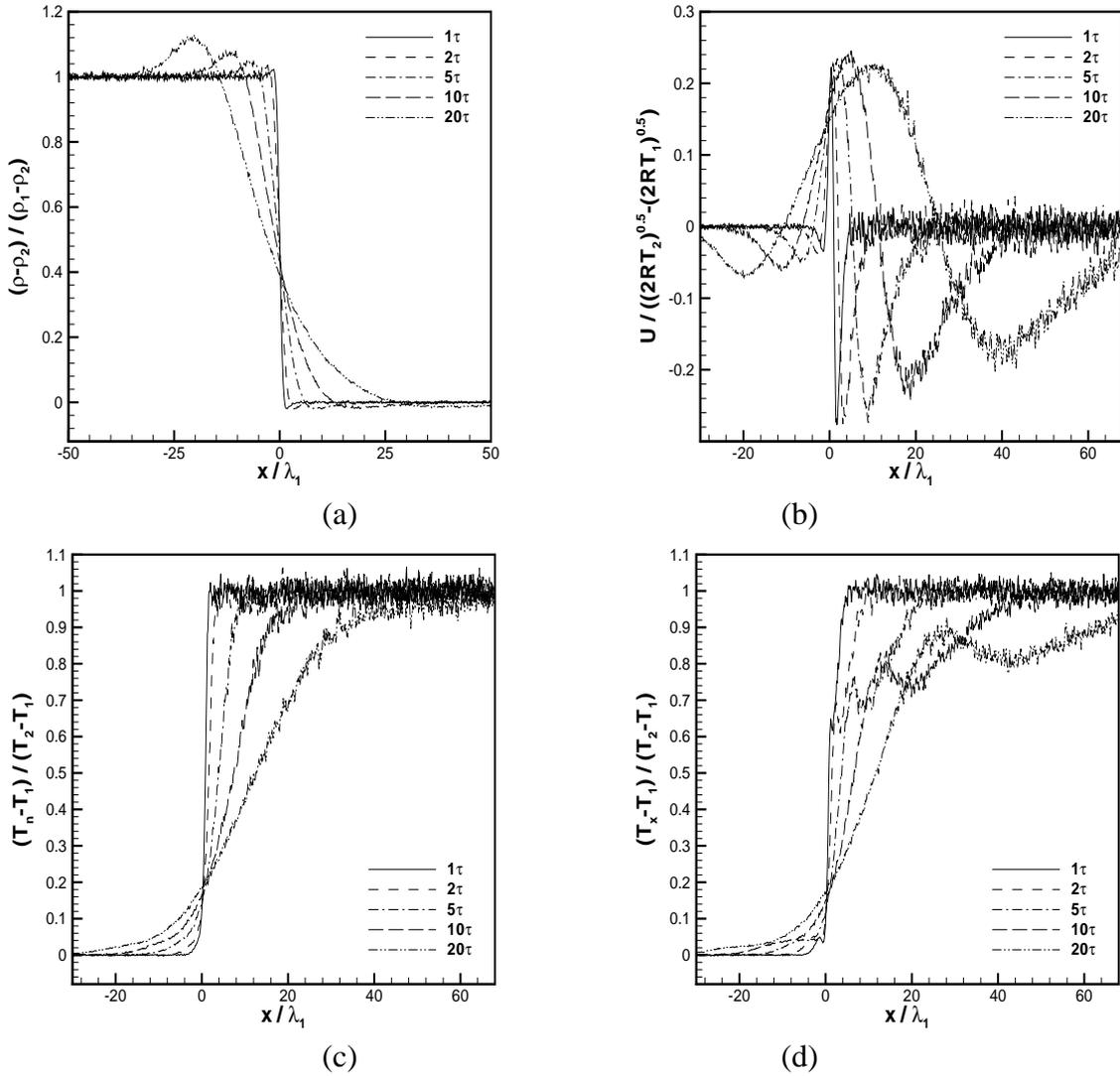

FIG. 3. DSMC results for contact discontinuities. (a) density; (b) flow velocity; (c) normal-component temperature; (d) parallel-component temperature.

In order to quantify the diffusion process, the thickness of contact discontinuity is defined as

$$d = \frac{x|_{\rho^*=0.2} - x|_{\rho^*=0.8}}{0.6} \quad (8)$$

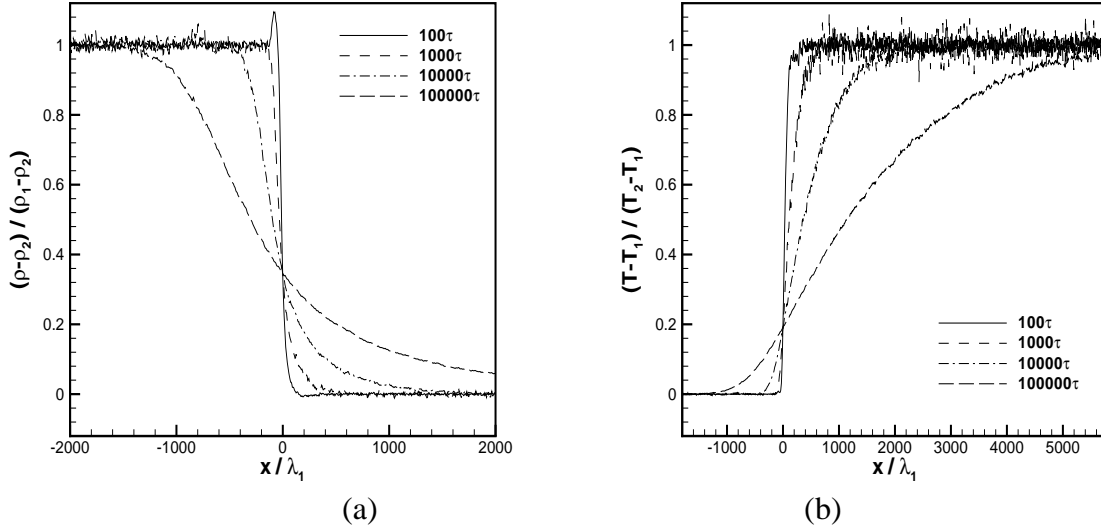

FIG. 4. DSMC results of contact discontinuities for longer time. (a) density; (b) temperature.

where $\rho^*$ is the normalized density as in Fig. 2a. The thickness may be defined as the inverse of the maximum gradient of the density profile. However, the error of the maximum gradient could be very large due to the statistical scatter in the DSMC results. Figure 5 shows the defined thickness of discontinuity for three values of strength at different times. It is clear that the thickness increases with the time and stronger discontinuities diffuse faster. At early time (less than 10 mean collision times), the thickness is proportional to the time. When the time is much larger, the thickness is proportional to the square root of the time, however. This is due to the fact that the non-equilibrium energy equations reduces to the heat transfer equation of the continuum flow for a large time, and the heat transfer equation rules that the heat transfer is proportional to the square root of the time. Based on Fig. 5, the thickness of a contact discontinuity in a sea level atmosphere is about 20μm when the strength is 1.1 and the time lasts for 6 μs, which means that the average diffusion speed is larger than 3m/s during this time span. In other words, contact discontinuities have obvious diffusion in viscous flows. This may imply that contact discontinuities should not be treated as jump discontinuities in certain cases.

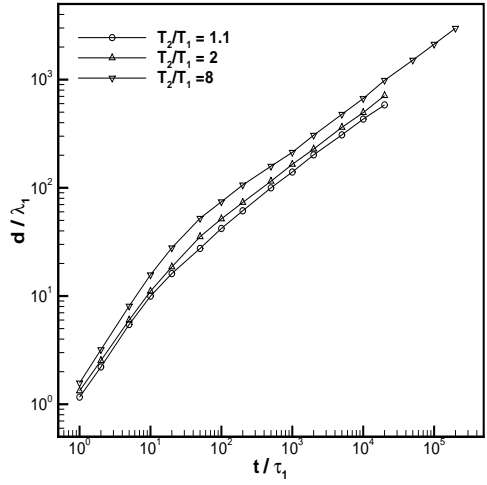

FIG. 5. Time evolution of the thickness of contact discontinuities.

Unlike contact discontinuity, the upstream and downstream conditions of a shock wave have to satisfy the Rankine-Hugoniot relations. For simplicity, the origin of the coordinate sits on the shock wave and monatomic gas argon is considered here. The upstream velocity should be determined from the shock wave Mach number $Ma_1$ and the downstream conditions follow the Rankine-Hugoniot relations:

$$u_1 = \sqrt{\frac{5}{6}} \frac{Ma_1}{\beta_1} \tag{9}$$

$$u_2 = u_1 \frac{Ma_1^2 + 3}{4 Ma_1^2} \tag{10}$$

$$T_2 = T_1 \frac{(5 Ma_1^2 - 1)(Ma_1^2 + 3)}{16 Ma_1^2} \tag{11}$$

$$\rho_2 = \rho_1 \frac{4 Ma_1^2}{Ma_1^2 + 3} \tag{12}$$

Using the same procedure for the contact discontinuity, the free molecular results for the shock wave are obtained as follows:

$$\frac{\rho(x,t)}{\rho_1} = \frac{1}{2}\left(1 - erf\left(\beta_1\left(\frac{x}{t} - u_1\right)\right)\right) + \frac{1}{2}\frac{n_2}{n_1}\left(1 + erf\left(\sqrt{\frac{T_1}{T_2}}\beta_1\left(\frac{x}{t} - u_2\right)\right)\right) \tag{13}$$

$$U(x,t) = \frac{\rho_1/\rho}{2\sqrt{\pi}\beta_1}e^{-\left(\beta_1\left(\frac{x}{t}-u_1\right)\right)^2} - \frac{\rho_2/\rho}{2\sqrt{\pi}\beta_2}e^{-\left(\beta_2\left(\frac{x}{t}-u_2\right)\right)^2}$$
$$+ \frac{u_1}{2}\frac{\rho_1}{\rho}\left(2 + erf\left(\beta_2\left(\frac{x}{t}-u_2\right)\right) - erf\left(\beta_1\left(\frac{x}{t}-u_1\right)\right)\right) \tag{14}$$

$$T_n(x,t) = \frac{T_1}{2}\frac{\rho_1}{\rho}\left(1 - erf\left(\beta_1\left(\frac{x}{t}-u_1\right)\right)\right) + \frac{T_2}{2}\frac{\rho_2}{\rho}\left(1 + erf\left(\beta_2\left(\frac{x}{t}-u_2\right)\right)\right) \tag{15}$$

$$T_x(x,t) = \frac{T_1}{\sqrt{\pi}} \frac{\rho_1}{\rho} \left( \beta_1 \left( \frac{x}{t} - u_1 \right) + 2\beta_1 u_1 \right) e^{-\left( \beta_1 \left( \frac{x}{t} - u_1 \right) \right)^2}$$

$$- \frac{T_2}{\sqrt{\pi}} \frac{\rho_2}{\rho} \left( \beta_2 \left( \frac{x}{t} - u_2 \right) + 2\beta_2 u_2 \right) e^{-\left( \beta_2 \left( \frac{x}{t} - u_2 \right) \right)^2}$$

$$+ \frac{T_1}{2} \frac{\rho_1}{\rho} \left( 1 + 2\beta_1^2 u_1^2 \right) \left( 1 - erf\left( \beta_1 \left( \frac{x}{t} - u_1 \right) \right) \right)$$

$$+ \frac{T_2}{2} \frac{\rho_2}{\rho} \left( 1 + 2\beta_2^2 u_2^2 \right) \left( 1 + erf\left( \beta_2 \left( \frac{x}{t} - u_2 \right) \right) \right) - \frac{U^2}{R}$$

(16)

These expressions involve the non-zero initial velocity, and profiles of shock waves based on these expressions are shown in Fig. 6. Clearly shock waves are diffused and the thickness of shock waves is about several mean free paths when the time is one mean collision time. The stronger shock wave, however, has larger thickness at an early time. An overshoot appears in the density profile on the higher density side as in the contact discontinuity case, but no undershoot is identified. The strength of the density overshoot in the shock wave is larger than that in the contact discontinuity for the same temperature difference between the two sides. The velocity profile is now scaled with the initial flow velocities instead of the speed of sound. It turns out that the velocity profile is very different from the contact discontinuity case and there is a local maximum on the lower density side. The normal component temperature is diffused smoothly across the shock. There are two obvious deflection points in the profile for the case with $T_2/T_1 = 8$. For the parallel or x component temperature, a large overshoot is observed in the profiles. The reason for the overshoot is because the fast moving molecules in the positive x-direction carries more energy from the higher velocity side than those from the opposite direction. Thus the overshoot increases with the shock strength.

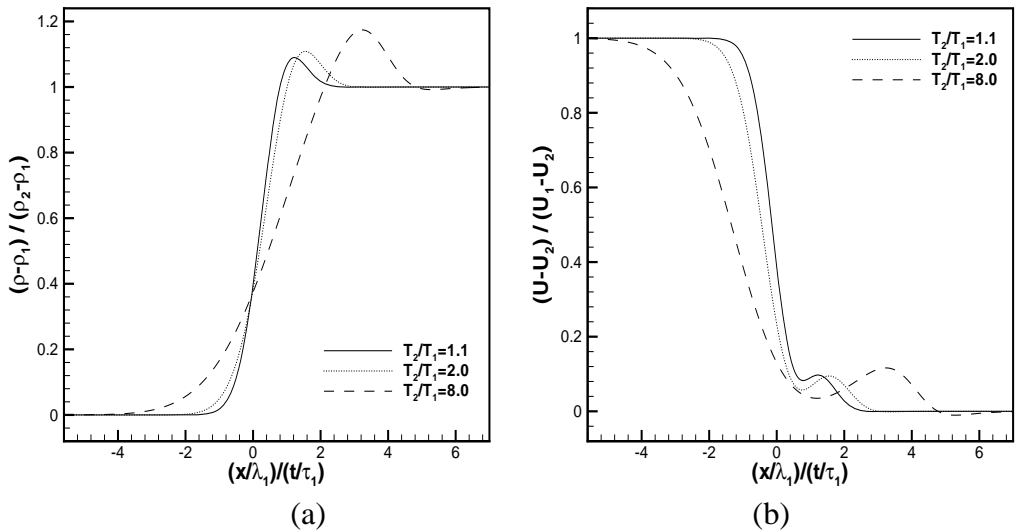

(a)  (b)

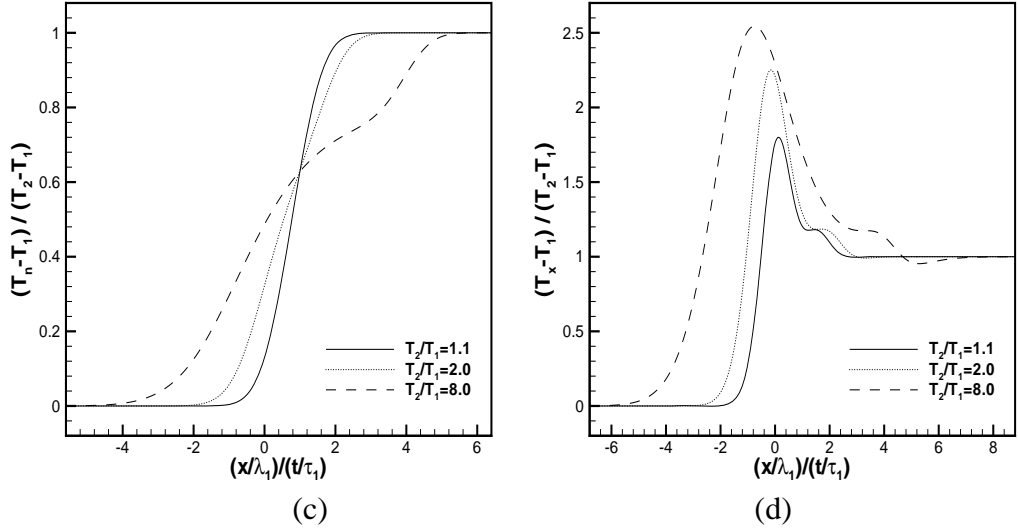

(c)   (d)

FIG. 6. Free molecular results of shock waves. (a) density; (b) flow velocity; (c) normal-component temperature; (d) parallel-component temperature.

Unlike the contact discontinuities, shock waves have stable structures. The jump discontinuity of a shock wave takes some time to reach its steady state. When the time is larger than the mean collision time, the free molecular results are not valid. The DSMC method is again employed to simulate the unsteady behavior of shock waves. Figure 7 shows the profiles of a shock wave with $T_2/T_1=8$ at different times (the corresponding shock Mach number is slightly less than 5). The time indicated in the plot looks awkward because the same time step used for the contact discontinuity is used for the shock wave cases. It turns out that the density overshoot decreases with the time and the density profile reaches a steady state later. Similarly, the temperature diffuses with the time and gradually reaches to the final shock structure. Notice that the overshoot in the x-component temperature decreases but does not disappear. Therefore, the parallel and normal components of the temperature do not reach equilibrium even at very large time. Figure 7d displays the total temperature profiles at different times.

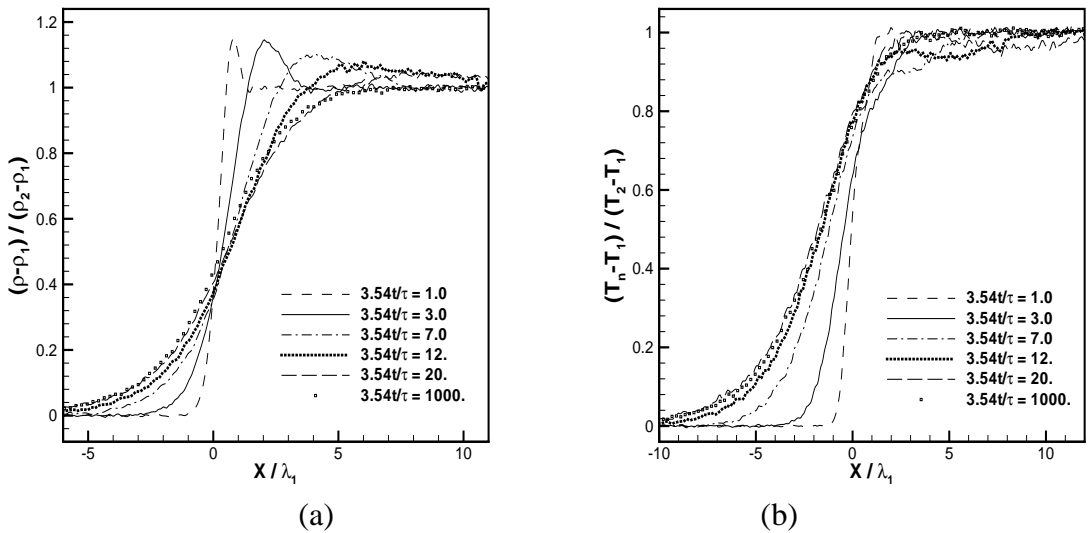

(a)   (b)

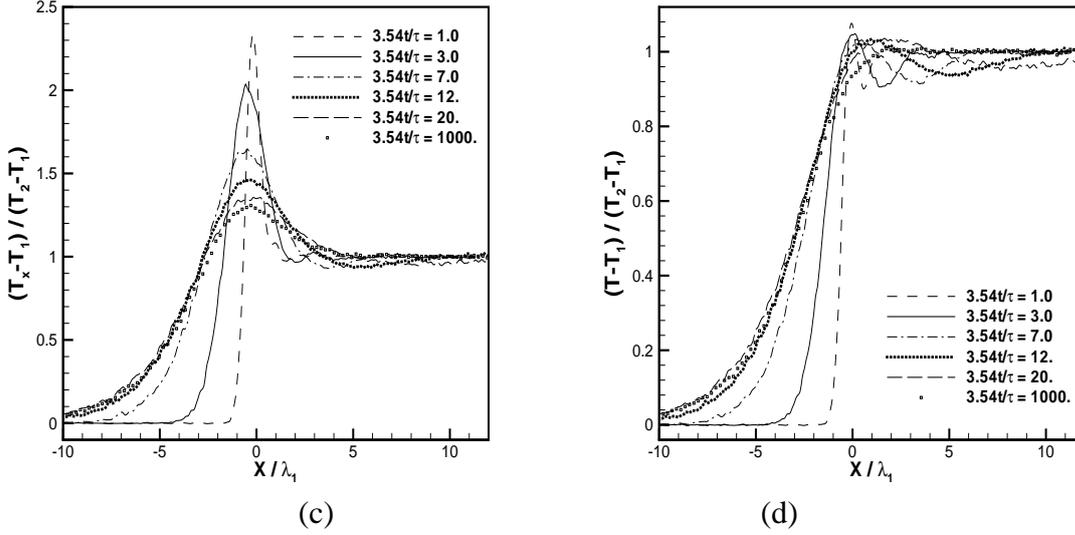

(c)                                (d)

FIG. 7. DSMC results for shock waves. (a) density; (b) normal-component temperature; (c) parallel-component temperature; (d) overall translational temperature.

    The time for a shock wave to reach its steady state can be estimated from the history of shock thickness. The shock thickness again is defined as in Eq. 8. In order to eliminate the effect of different definition, the shock thickness is normalized by its thickness at the steady state. Figure 8 shows the evolution of the normalized shock thickness for two shocks. It is found that the shock thickness initially increases with time, reaches a maximum, and has a slight decrease to approach the final thickness. For a Mach 5 shock wave, it only takes about 10 mean collision times to reach a steady state. It takes more time for a weaker shock wave to develop, however. For instance, it will take more than 1000 mean collision times for a Mach 1.1 shock wave to develop. There are two reasons for the slow evolution of a weak shock wave. One is that the nonlinear advection term in the fluid equations approaches to the linear limit as shock becomes weak, and the lack of steeping mechanism make the thickness large. Navier-Stokes results predict that shock thickness approaches infinite when the shock strength gets weaker and weaker. The other is because the diffusion speed decreases when the shock strength decreases.

    The above solution from discontinuities of contact and shock waves give us a clear picture about the gas evolution, from the collision-less solution to the construction of the NS solution due to intensive particle collisions. The Euler solution can be only reached on a time scale which is much larger than the particle collision time. For the hypersonic flow computations, with the scale of mesh size, the particles may not encounter enough collisions to form equilibrium wave structures. A reliable numerical scheme should somehow respect the physical process presented in this section.

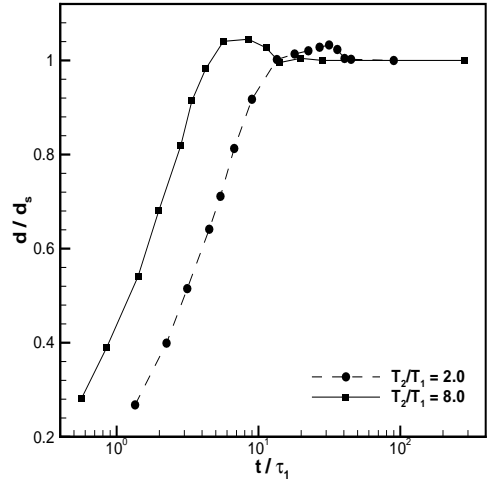

FIG. 8. Time evolution of shock wave thickness.

## IV. PHYSICAL PROCESS IN THE GAS-KINETIC SCHEME

Instead of using the macroscopic Euler and NS equations, the gas-kinetic scheme (GKS) is based on the Boltzmann equation [8], which is far more general than the NS equations in the description of flow physics. The gas-kinetic BGK scheme is based on the kinetic BGK model

$$f_t + u f_x = (g - f)/\tau ,$$

where $f$ is the gas distribution function and $g$ is the equilibrium state. For the finite volume gas-kinetic scheme (GKS), the integral solution of the above BGK model is used for the flux evaluation,

$$f = \frac{1}{\tau}\int_0^t g(x',t',u) e^{-(t-t')/\tau} dt' + e^{-t/\tau} f_0(x - ut) = (1 - e^{-t/\tau}) f_{II} + e^{-t/\tau} f_I(x - ut) ,$$

where $f_I$ is the initial gas distribution function constructed based on the discontinuous initial data, and its solution represents a free transport process along particle trajectory. The term $f_{II}$ is related to the integration of equilibrium state, which accounts for particle collisions. In the limit of $t \gg \tau$, the term $f_{II}$ automatically gives a distribution function which recovers the NS solution. The Euler solution can be considered as a limiting solution of the NS one. Therefore, in the GKS the basic physical process underlying $f_I$ is the same as the FVS method, and underlying $f_{II}$ it gives a process to go to the NS solution. The FDS solution can be considered as the limiting solution of $f_{II}$. Therefore, the GKS method is a unification of upwind FVS and FDS schemes, and it provides an evolution model from FVS to FDS. This model is consistent with the real physical process presented in the last section.

On the other hand, in the GKS scheme a piecewise discontinuous initial data is usually used, and the upwind property is intrinsically rooted in its free transport term $f_I$. However, the $f_{II}$ term represents the drifting of the equilibrium state, which recovers a continuous distribution across a cell interface. So, this term is similar to the central difference method. If a continuous initial reconstruction is used in the smooth flow regime, the upwind property in $f_I$ disappears automatically. With the combination of $f_{II}$ term, the GKS method goes back to the traditional Lax-Wendroff type central difference scheme for the NS equations. Therefore, the use of upwind or central difference depends solely on the smoothness of the initial reconstruction in the GKS method. It doesn't settle down to the upwind or central difference schools from the starting point in the design of a numerical scheme. In other words, the GKS scheme is a unification of the central difference and upwind methods. Also, the traditional way of extending upwind differencing to multidimensional equations is by doing it dimension by dimension. This means that numerically all transport is done by waves moving normal to the cell faces. This makes the upwind scheme be sensitive in the shock solution which depends on the relative orientation between the shock front and mesh lines. However, for the GKS scheme [21], a natural multi-dimensional solution can be obtained from an initial piecewise continuous flow distribution with variation in both normal and tangential directions of a cell interface. In smooth flow region, the GKS goes to multi-dimensional central-difference methods.

The GKS method unifies the approach of the FVS and FDS fluxes, and the upwind and central difference discretization. The weighting function between FVS and FDS depends on a relaxation process from $f_I$ to $f_{II}$. This evolution process is consistent with the physical one from a discontinuity presented in the last section. However, the quantitative dissipation in the GKS method is closely related to the discontinuous jump. In other words, the added dissipation is related to the cell resolution. In comparison with a purely FVS scheme, the advantage of the GKS method is that this amount of dissipation is controlled and reduced through the relaxation to equilibrium state, and the relaxation depends on the physical situation. In the high Reynolds number flow, such as the boundary layer, due to the big ratio between the numerical time step and the particle collision time, the NS flux is automatically obtained in the GKS scheme due to the dominant of $f_{II}$ term in the final distribution function $f$ at a cell interface. However, in the strong shock layer, especially in the high Mach number case, the equivalence between the particle collision time and the numerical time step provides enough dissipation from $f_I$ for the construction of a stable shock structure. Fig. 9 presents the high speed flow simulation using GKS at M=20 and 30 at different incident angle around a circular cylinder, where carbuncle phenomenon has never been observed, even up to M=100. In summary, the shock structure obtained in the GKS scheme is not a purely numerical one, it is constructed through a valid non-equilibrium physical process of particle transport and collision, even though the scale of shock thickness is numerically enlarged to the scale of mesh size.

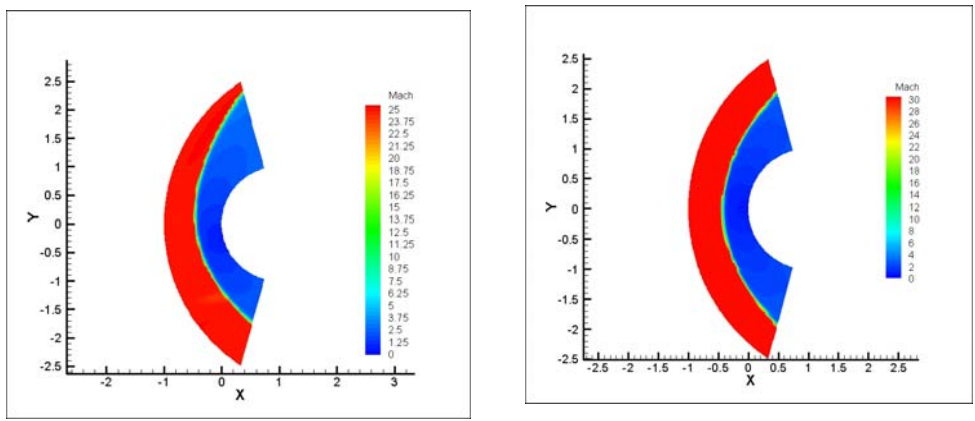

FIG. 9. Flow around half cylinder. M=20 with incident angle $15^o$ (left), and M=30 normal shock (right).

# V. CONCLUDING REMARKS

The valid time evolution from a discontinuity is studied in this paper through kinetic and DSMC methods. More specifically, the time evolution of contact discontinuities and shock waves in a non-equilibrium flow situation has been obtained. The physical process from a discontinuity starts from the collision-less Boltzmann solution to the dissipative Navier-Stokes wave construction. The benchmark solutions presented in this paper can be used for the mathematical modeling of non-equilibrium flows in the construction of reliable CFD methods. Even though the shock capturing scheme in CFD may not use the precise process of non-equilibrium flow physics on a microscopic scale, to use a similar numerical representation of non-equilibrium evolution model seems necessary in the construction of robust and accurate shock capturing schemes. The underlying physical pictures of flux vector splitting (FVS) and flux difference splitting (FDS) are two limiting cases of the above non-equilibrium gas evolution model. The direct use of exact Riemann solver in FDS scheme triggers shock instability, such as the carbuncle phenomenon, because the FDS is a limiting solution with the assumption of infinite number of particle collisions which cannot be satisfied in a numerical shock layer. Therefore, the particle free transport and collision mechanism has to be taken into account in the numerical scheme. The basic assumption of infinite number of particle collision to form equilibrium state instantaneously in the Godunov method is valid only in the cases of smooth flow region or inside weak numerical shock layers. On the other hand, the gas-kinetic scheme (GKS) uses a non-equilibrium gas evolution model in its numerical flux construction, where the particle free transport, collision, and the formation of equilibrium state have been followed. The GKS unifies the FVS and FDS methods and the transition from one to the other depends on the ratio between the time step and particle collision time. At the same time, the GKS unifies the central difference and upwind schemes. The distinction between the above two approaches is solely based on the reconstruction of the initial data. In the smooth flow regime, the discontinuity at a cell interface disappears and the GKS goes back to the Lax-Wendroff type central difference scheme for the NS solution.

In summary, the Godunov method uses equilibrium flow solution with the assumption of infinite number of particle collisions in its numerical flux construction. This mechanism is inconsistent with the flow physics inside a highly non-equilibrium numerical shock layer. The persisting shock instability in most shock capturing schemes clearly indicates that the use of exact Euler solution in the construction of numerical flux is problematic even though they are extremely successful in low and modest flow speed.

**ACKNOWLEDGMENTS**
This work was supported by Hong Kong Research Grant Council 621709 and RPC10SC11, National Natural Science Foundation of China (Project No. 10928205, 50836007，90816012), National Key Basic Research Program (2009CB724101).